 \documentclass[10pt, technote]{IEEEtran} 
\usepackage{cite}
\usepackage[dvips]{graphicx}
\ifCLASSINFOpdf
\else
\fi
\usepackage[cmex10]{amsmath}
\ifCLASSOPTIONcompsoc
  \usepackage[caption=false,font=normalsize,labelfont=sf,textfont=sf]{subfig}
\else
  \usepackage[caption=false,font=footnotesize]{subfig}
\fi

\usepackage{mathtools}
\usepackage{xcolor}

\begin{document}
%
\title{Electromagnetic Fields Radiated\\by a Circular Loop with Arbitrary Current}
%
%
%

\author{Mohamed~A.~Salem
        and~Christophe~Caloz,~\IEEEmembership{Fellow,~IEEE}
\thanks{M.~A.~Salem and C.~Caloz are with the Poly-Grames Research Center, \'{E}cole Polytechnique de Montr\'{e}al, Montr\'{e}al, Qu\'{e}bec, H3T 1J4 Canada e-mail: (see http://www.calozgroup.org/members.html).}}

\maketitle

\begin{abstract}
We present a rigorous approach to compute the electromagnetic fields radiated by a thin circular loop with arbitrary current. We employ a polar transmission representation along with a Kontorovich-Lebedev transform to derive integral representations of the field in the interior and exterior regions of a sphere circumscribing the loop. The convergence of the obtained expressions is discussed and comparisons with full-wave simulation and other methods are shown.
\end{abstract}


%
\IEEEpeerreviewmaketitle

\section{Introduction}
%
%
%
%
Electromagnatic (EM) radiation by a circular current loop is extensively investigated in the literature (see e.g.~\cite{Li:97,Hamed:13} and references therein, and comments thereon~\cite{Werner:01,Gaffoor:01,Zheng:14,Hamed:14}). However, the majority of the studies investigated the far field pattern. The evaluation of the near field in closed form is generally limited to constant or cosine current distributions due to difficulties in evaluating the corresponding integrals.

Here, we present a rigorous approach to express the electromagnetic fields radiated by a thin current loop excited with an arbitrary current form. The presented approach employs the Kontorovich-Lebedev (KL) transform and the Fourier series expansion to express the radiated field in the interior and exterior regions of a sphere circumscribing the loop. The resulting expression includes an integral whose convergence is detailed. An alternative representation in terms of a residue series sum is presented for field evaluation in the interior region. The field expressions could thus be evaluated in closed form or by numerical integration techniques with ease, and they do not suffer from any irregularities or artificial discontinuities in contrast to previously reported approaches~\cite{Li:97,Hamed:13}.

\begin{figure}[!t]
\centering
\includegraphics[width=3.25in]{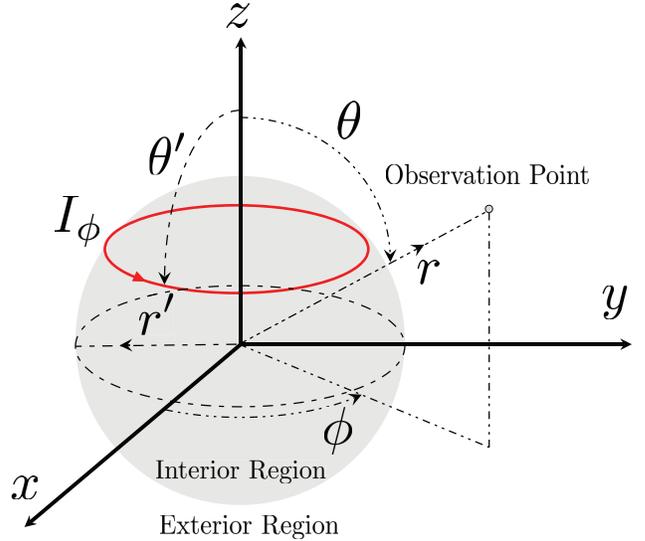}
\caption{Geometry of the circular current loop.}
\label{fig:01}
\end{figure}

\section{Formulation of the Problem}
The geometry of the problem is shown in Fig.~\ref{fig:01}, where a circular current loop is placed in free-space and carries an arbitrary electric current along its circumference. In spherical coordinates, $(r,\theta,\phi)$, the loop is located at $r = r'$ and $\theta=\theta'$. The electric current density on the loop is given by%
\begin{equation}
\label{eq:J}
\mathbf{J}\left(\mathbf{r}\right) = \frac{I\left(\phi\right)}{r'} \delta\left(r-r'\right)\delta\left(\theta-\theta'\right)\mathbf{\hat{\phi}},
\end{equation}%
where $I(\phi)$ is the current distribution in the azimuthal direction and $\delta(r)$ is the Dirac delta. An observer at $(r,\theta,\phi)$ is said to be in the interior region when $r<r'$ and in the exterior region otherwise. The time-harmonic dependence of $\exp(-i\omega t)$ is assumed throughout.

The electric and magnetic fields are expressed in terms of Hertz potentials, $\Pi_e$ and $\Pi_h$, oriented along $\mathbf{\hat{r}}$, viz.%
\begin{equation}
\label{eq:field}
\begin{split}
\mathbf{E} &= \nabla \times \nabla \times \mathbf{\hat{r}} \Pi_e + i \omega \mu \nabla \times \mathbf{\hat{r}} \Pi_h, \\
\mathbf{H} &= - i \omega \epsilon \nabla \times \mathbf{\hat{r}} \Pi_e + \nabla \times \nabla \times \mathbf{\hat{r}} \Pi_h,
\end{split}
\end{equation}%
where $\epsilon$ and $\mu$ are the free-space permittivity and permeability, respectively. The choice of the Hertz potential orientation allows for the separation of the fields into transverse electric (TE) and transverse magnetic (TM) components with respect to $r$~\cite{Morse:53}. It thus follows that the TE components are derived from $\Pi_e$, while the TM components are derived from $\Pi_h$.

The Hertz potentials in~\eqref{eq:field} may be expressed in terms of a single auxiliary Green's function for transverse (with respect to $r$) current elements~\cite[pp.~224--225]{Felsen:94} as%
\begin{equation}
\label{eq:hertz}
\begin{split}
r\Pi_e &= \frac{i}{\omega\epsilon} \int_{-\pi}^{\pi} \nabla' \times \nabla' \times\mathbf{\hat{r}'}\mathcal{G} \cdot \mathbf{\hat{\phi}'} I\left(\phi'\right) d \phi', \\
r\Pi_h &= \int_{-\pi}^{\pi} \nabla' \times\mathbf{\hat{r}'}\mathcal{G} \cdot \mathbf{\hat{\phi}'} I\left(\phi'\right) d \phi',
\end{split}
\end{equation}%
where the prime denotes operation on the source coordinates. The auxiliary Green's function satisfies the relation%
\begin{equation}
\label{eq:aux_green}
\left( \nabla^2 +k^2 \right) \nabla_{\perp}^2 \mathcal{G}\left(\mathbf{r};\mathbf{r}'\right) = \delta\left(\mathbf{r}-\mathbf{r}'\right),
\end{equation}
with $k=\omega/\sqrt{\epsilon \mu}$ being the wave number, $\nabla_{\perp}^2$ the transverse Laplacian with respect to $r$, and $\mathbf{r}$ and $\mathbf{r}'$ the position vectors of the observer and the source, respectively. Nevertheless, we will solve for the conventional Green's function that satisfies%
\begin{equation}
\label{eq:green}
\left( \nabla^2 +k^2 \right) G\left(\mathbf{r};\mathbf{r}'\right) = - \delta\left(\mathbf{r}-\mathbf{r}'\right),
\end{equation}%
then deduce $\mathcal{G}$ from $G$ using the relation%
\begin{equation}
\label{eq:green_aux}
-r'^2 \nabla'_\perp \cdot \prescript{}{\perp}{\nabla'} \mathcal{G}\left(\mathbf{r};\mathbf{r}'\right) = r r' G\left(\mathbf{r};\mathbf{r}'\right).
\end{equation}

Note that~\eqref{eq:aux_green} and~\eqref{eq:green} are defined for an infinitesimal current element located at $\mathbf{r}'$. In what follows, we compute $G$ for an element located at $(r',\theta',\phi')$, then derive the corresponding $\mathcal{G}$ for the current loop via~\eqref{eq:green_aux}. Next, we express~\eqref{eq:green} in terms of a polar transmission-line (along $\theta$) with eigenfunctions evaluated in the cross section transverse to $\theta$. This representation is different from the conventional notion of transmission along $r$ that is adopted in~\cite[pp.~699--701]{Felsen:94},~\cite{Tai:94}. This polar transmission-line representation avoids unphysical field discontinuities that may occur from uncareful formulation of the dyadic Green's function and is more natural to the geometry of the problem than radial transmission-line representation.

Rewriting~\eqref{eq:green} in explicit form as%
\begin{equation}
\label{eq:green_sph}
\begin{split}
\left( \nabla^2 +k^2 \right) G\left(r,\theta,\phi;r',\theta',\phi'\right) = \\
-\frac{\delta\left(r-r'\right)\delta\left(\theta-\theta'\right)\delta\left(\phi-\phi'\right)}{r^2 \sin\left(\theta'\right)},
\end{split}
\end{equation}
one can represent the azimuthal dependence of the Green's function in terms of a Fourier series as%
\begin{equation}
\label{eq:FS}
G\left(r,\theta,\phi\right) = \sum_{m=-\infty}^{\infty}{\tilde{g}_m\left(r,\theta\right) e^{i m \phi}}.
\end{equation}%
Following~\cite{Salem:12}, the radial dependence is represented in terms of a KL transform, where the KL transform integral pair reads~\cite{Jones:80}%
\begin{align}
\tilde{g}_m\left(r,\theta\right) &= \frac{k}{\pi} \int_{-i\infty}^{i\infty}{g_m\left(\nu,\theta\right) j_{\nu-\frac{1}{2}}\left(k r\right) \nu d \nu}, \label{eq:iKL} \\
g_m\left(\nu,\theta\right) &= \int_{0}^{\infty}{\tilde{g}_m\left(r,\theta\right) h_{\nu-\frac{1}{2}}^{(1)}\left(k r\right) d r}, \nonumber
\end{align}
where $ j_{\nu}(k r)$ and $h_{\nu}^{(1)}(k r)$ are the spherical Bessel and Hankel functions of first kind and order $\nu$. 

Inserting~\eqref{eq:iKL} into~\eqref{eq:FS} and substituting the result into~\eqref{eq:green_sph}, then using the orthogonality properties of the exponential functions and the Hankel function~\cite{Jones:80} yields the polar ordinary differential equation%
\begin{equation}
\label{eq:green_1d}
\begin{split}
\left( \frac{1}{\sin\left(\theta\right)} \frac{d}{d \theta} \left[ \sin\left(\theta\right) \frac{d}{d \theta} \right] + \nu^2 - \frac{1}{4} \right) g_m\left(\nu,\theta\right) = \\
 -e^{-i m \phi'} h_{\nu-\frac{1}{2}}^{(1)}\left(k r'\right) \frac{\delta\left(\theta-\theta'\right)}{\sin\left(\theta'\right)}.
\end{split}
\end{equation}
The solution of~\eqref{eq:green_1d} is given in terms of associated Legendre functions and must be bounded at $\theta = 0,\pi$, hence~\cite{Salem:12}%
\begin{equation}
\label{eq:gm}
\begin{split}
g_m\left(\nu,\theta\right) = -\frac{\pi}{2} \frac{e^{-i m \phi'} h_{\nu-\frac{1}{2}}^{(1)}\left(k r'\right)}{\sin\left(\pi\left[\nu - \frac{1}{2} - m\right]\right)} \\
\times \frac{\Gamma\left(\nu + m - \frac{1}{2}\right)}{\Gamma\left(\nu - m - \frac{1}{2}\right)} P_{\nu - \frac{1}{2}}^{-m}\left(\cos\left(\theta_<\right)\right) P_{\nu - \frac{1}{2}}^{-m}\left(-\cos\left(\theta_>\right)\right),
\end{split}
\end{equation}%
where $\Gamma(z)$ is the gamma function and $\theta_>|\theta_<$ is the greater$|$lesser of $\theta$ and $\theta'$.

The relation~\eqref{eq:green_aux}, after the application of the eigenfunction expansion to the radial and azimuthal dependencies, reduces to%
\begin{equation}
\label{eq:aux_green_1d}
\gamma_m\left(\nu,\theta\right) = - \frac{g_m\left(\nu,\theta\right)}{\nu\left[\nu+1\right]}.
\end{equation}%
Substituting~\eqref{eq:aux_green_1d} into~\eqref{eq:iKL} then back into~\eqref{eq:FS} gives the polar transmission-line representation of the auxiliary Green's function as%
\begin{equation}
\label{eq:1d_green_int}
\begin{split}
\mathcal{G}\left(r,\theta,\phi;r',\theta',\phi'\right) = \\ \frac{k}{\pi} \sum_{m=-\infty}^{\infty}{\int_{-i\infty}^{i\infty}{ \gamma_m\left(\nu,\theta\right) e^{i m \phi} j_{\nu-\frac{1}{2}}\left(k r\right)  \nu d \nu}}.
\end{split}
\end{equation}
We should note here that the sum over $m$ from $-\infty$ to $\infty$ is formal, since in practice the azimuthal exponentials are combined into a cosine function and the summation limits are changed in fashion similar to~\eqref{eq:1d_green_srs}, since $P_{\nu-1/2}^{-m}(\cos(\theta))$ and $P_{\nu-1/2}^{-m}(-\cos(\theta))$ are respectively bounded at $\theta = 0,\pi$ if $\mathrm{Re}\{m\} > 0$.

Next, we look into the asymptotic behavior of~\eqref{eq:1d_green_int} to investigate the existence of the integral. From~\cite[pp.~713--715]{Felsen:94}, the exponential asymptotic behavior of Bessel functions is%
\begin{equation}
J_\nu\left(z\right), H_\nu^{(1)}\left(z\right) \sim e^{|\nu| \frac{\pi}{2}} / |\nu|^{\frac{1}{2}},\; \nu\rightarrow \pm i \infty, \nonumber
\end{equation}%
with the relation between the spherical Bessel (or Hankel) function and its cylindrical counterpart is given by%
\begin{equation}
z_\nu\left(k r\right) = \sqrt{\frac{\pi}{2 k r}} Z_{\nu+\frac{1}{2}}\left(k r\right).\nonumber
\end{equation}
The exponential asymptotic behavior of associated Legendre functions is given by~\cite[(8.721)]{Gradshteyn:07}%
\begin{equation}
P_{\nu-\frac{1}{2}}^{-m}\left(\pm\cos\left(\theta\right)\right) \sim e^{\pm |\nu| \theta} / |\nu|^{\frac{1}{2}\mp  m},\; \nu\rightarrow \pm i \infty. \nonumber
\end{equation}%
We infer that the integrand in~\eqref{eq:1d_green_int} decays exponentially as%
\begin{equation}
e^{-|\nu| |\theta - \theta'|} /|\nu|^2, \; \nu\rightarrow \pm i \infty. \nonumber
\end{equation}
and the integral in~\eqref{eq:1d_green_int} exists. However, instead of directly evaluating the integral, we can make use of Cauchy residue theorem by closing the integration contour in the right hand side of the complex $\nu$-plane and collecting the residue contributions. We note from~\eqref{eq:gm} that the integrand has simple poles at%
\begin{equation}
\nu_n = n + m + \frac{1}{2}, \; n=0,1,2,\dots \nonumber
\end{equation}%
The residue series sum representation of the integral is thus given by%
\begin{equation}
\label{eq:1d_green_srs}
\begin{split}
\mathcal{G}\left(r,\theta,\phi;r',\theta',\phi'\right) = i \pi k \\ 
\times \sum_{m=0}^{\infty}{\sum_{n=m}^{\infty}{ \varepsilon_m \cos\left( m \left[\phi-\phi'\right]\right) \frac{h_{n}^{(1)}\left(k r'\right) j_{n}\left(k r\right) }{\left[n + \frac{3}{2}\right] }}} \\
\times \frac{\left(n-m\right)!}{\left(n+m\right)!} P_{n}^{m}\left(\cos\left(\theta\right)\right) P_{n}^{m}\left(\cos\left(\theta'\right)\right),
\end{split}
\end{equation}%
with $\varepsilon_m = 1$, $m=0$ and $\varepsilon = 2$ otherwise, $P_n^m(z)$ is the conventional associated Legendre polynomial and we have made use of their negative argument property~\cite[(8.736)]{Gradshteyn:07} $P_n^m(z) = -1^n P_n^m(-z)$. From~\cite[pp.~713--715]{Felsen:94}, the asymptotic exponential behavior of the Bessel functions is given by%
\begin{align}
J_n\left(z\right) &\sim \sqrt{\frac{1}{2\pi n}}\left[\frac{2 n}{e z} \right]^{-n}, \; & n \rightarrow \infty, \nonumber \\
H_n^{(1)}\left(z\right) &\sim \sqrt{\frac{1}{2\pi n}}\left[\frac{2 n}{e z} \right]^{n}, \; & n \rightarrow \infty, \nonumber
\end{align}%
thus the dominant behavior of the series summand reduces to%
\begin{equation}
\left[\frac{r}{r'} \right]^{n}, \; n \rightarrow \infty. \nonumber
\end{equation}%
This behavior suggests that~\eqref{eq:1d_green_srs} is appropriate for field evaluation in the interior region, while field evaluation in the exterior region requires carrying out the integral in~\eqref{eq:1d_green_int}.

To evaluate the field, the Hertz potentials are first computed from the auxiliary Green's functions using~\eqref{eq:hertz} and then substituted into~\eqref{eq:field}. It should be noted here, that the additional integral in~\eqref{eq:hertz} reduces to a Fourier series expansion of $I(\phi)$. This completes the formulation of the problem of EM radiation by a circular current loop.


\section{Evaluation of the Fields}
The expressions for the Hertz potentials are obtained by substituting~\eqref{eq:1d_green_int} into~\eqref{eq:hertz}. Inserting the Hertz potential expressions into~\eqref{eq:field} yields the expressions for the electromagnetic fields. These expressions are given in the appendix.

Next, we evaluate the field for a time-harmonic excitation corresponding to a free-space wavelength $\lambda_0 = 60\,\mathrm{mm}$ and wave number $k_0 = 2\pi/\lambda_0$. The current loop is located at $\theta'=\pi/2$ and $r'=20\,\mathrm{mm}$ $(\lambda_0/3)$ and is carrying a current distribution $I(\phi) = \exp(-\phi/(2\pi))\exp(i\phi)$, which is chosen to model one of the current modes on a circular leaky-wave antenna~\cite{Al-Bassam:14}. The fields are computed in the Cartesian plane $z = 10 \,\mathrm{mm}$ $(\lambda_0/6)$ for $|x| =20 \,\mathrm{mm}$ $(2\lambda_0/3)$ and $y=0$. In our proposed approach, the fields are computed using~\eqref{eq:1d_green_srs} in the interior region up to $r/r' = 0.85$ and using~\eqref{eq:1d_green_int} elsewhere. The azimuthal series is truncated after 16 terms and the radial series in~\eqref{eq:1d_green_srs} is truncated after 32 terms when employed. The integral in~\eqref{eq:1d_green_int} is computed using Simpson's rule and is truncated at $|\nu| = 18$ when employed. For comparison, full-wave simulation using HFSS by ANSYS~\cite{HFSS:13} and results from~\cite{Li:97} are considered. We note that the results from~\cite{Li:97} and~\cite{Hamed:13} are identical, however we chose to use the formulation of~\cite{Li:97} due to its compact representation. For the full-wave simulation, the Cartesian computation domain is set to $120 \,\mathrm{mm} \times 120 \,\mathrm{mm} \times 40 \,\mathrm{mm}$ $(2\lambda_0 \times 2\lambda_0 \times 2\lambda_0/3)$ and is centered at the origin of the coordinate system. The current loop is emulated by 36 azimuthally oriented dipoles uniformly distributed around the loop circumference and the magnitude and phase of their currents are individually set. To facilitate the computation of the fields, a dielectric sphere of radius $r_{\mathrm{sphere}} = 5 \,\mathrm{mm}$ $(\lambda_0/12)$ with relative permittivity $\epsilon_{r,\mathrm{sphere}} = 1.001$ is placed at the center of the computation domain. To obtain the results using the approach in~\cite{Li:97}, the azimuthal and the polar series are truncated after 16 terms each for the same computation parameters used in our approach.

\begin{figure}[!t]
\centering
\subfloat[]{\includegraphics[width=3in]{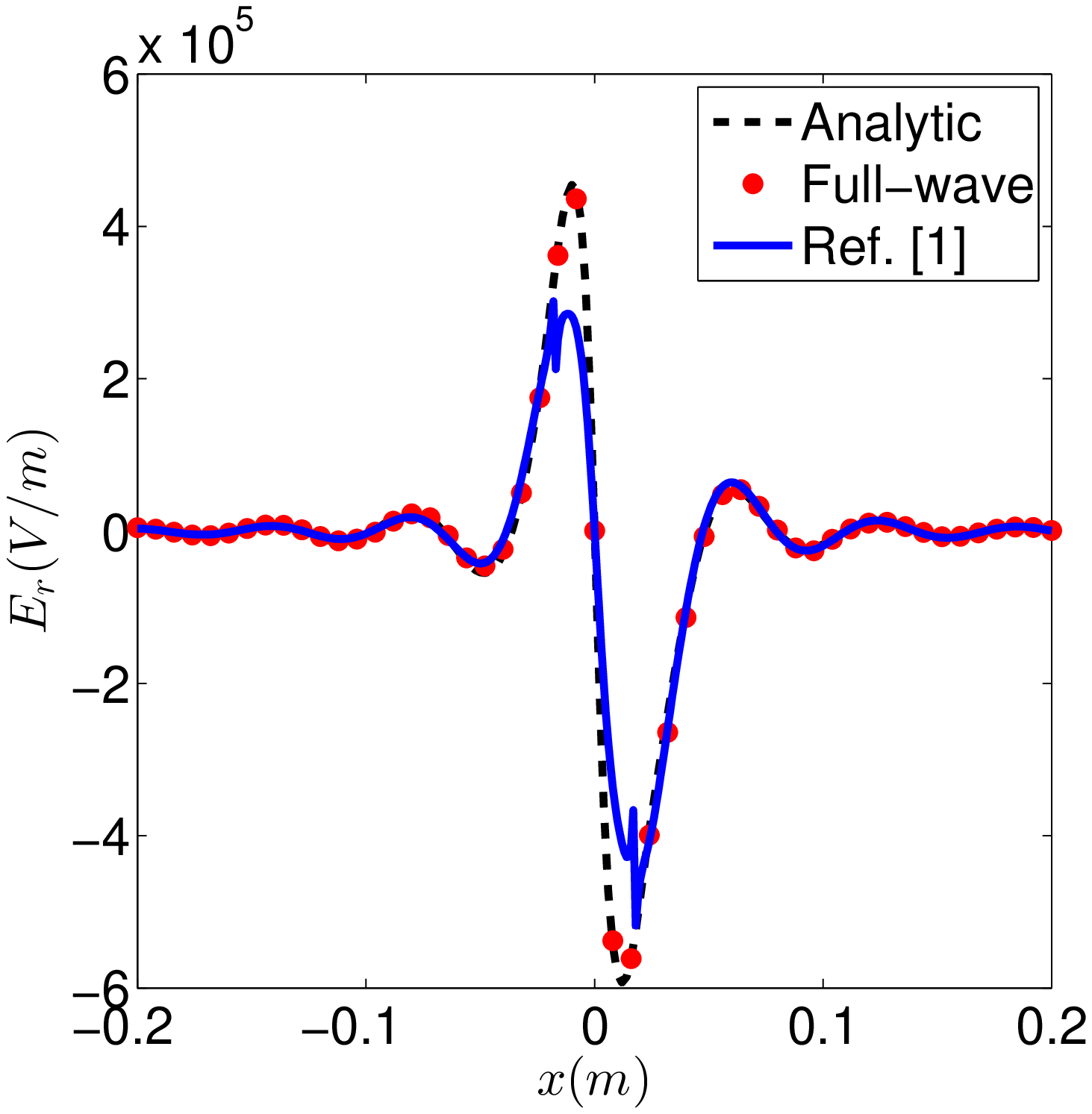}%
\label{fig:2a}}
\\
\subfloat[]{\includegraphics[width=3in]{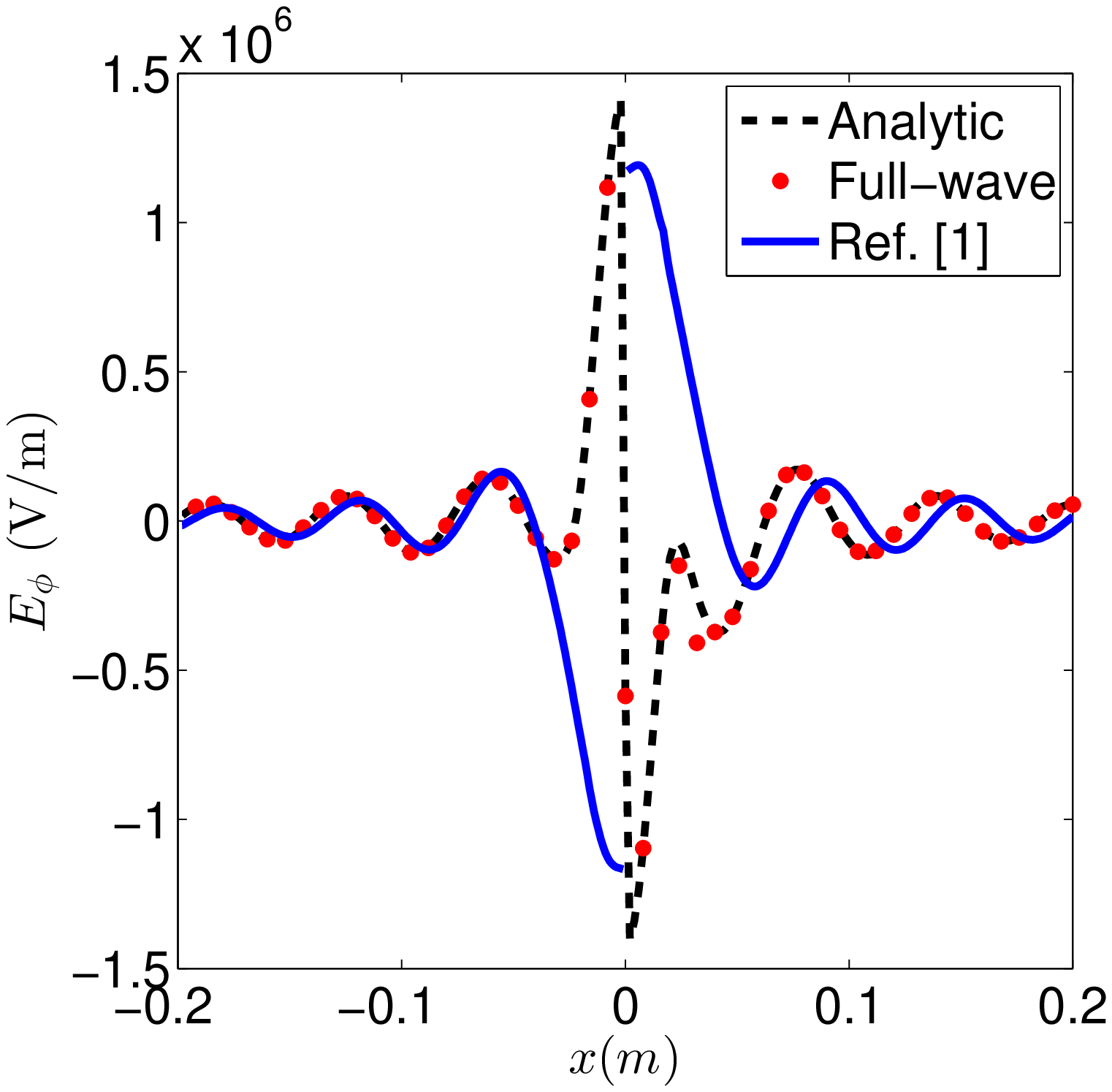}%
\label{fig:2b}}
\\
\subfloat[]{\includegraphics[width=3in]{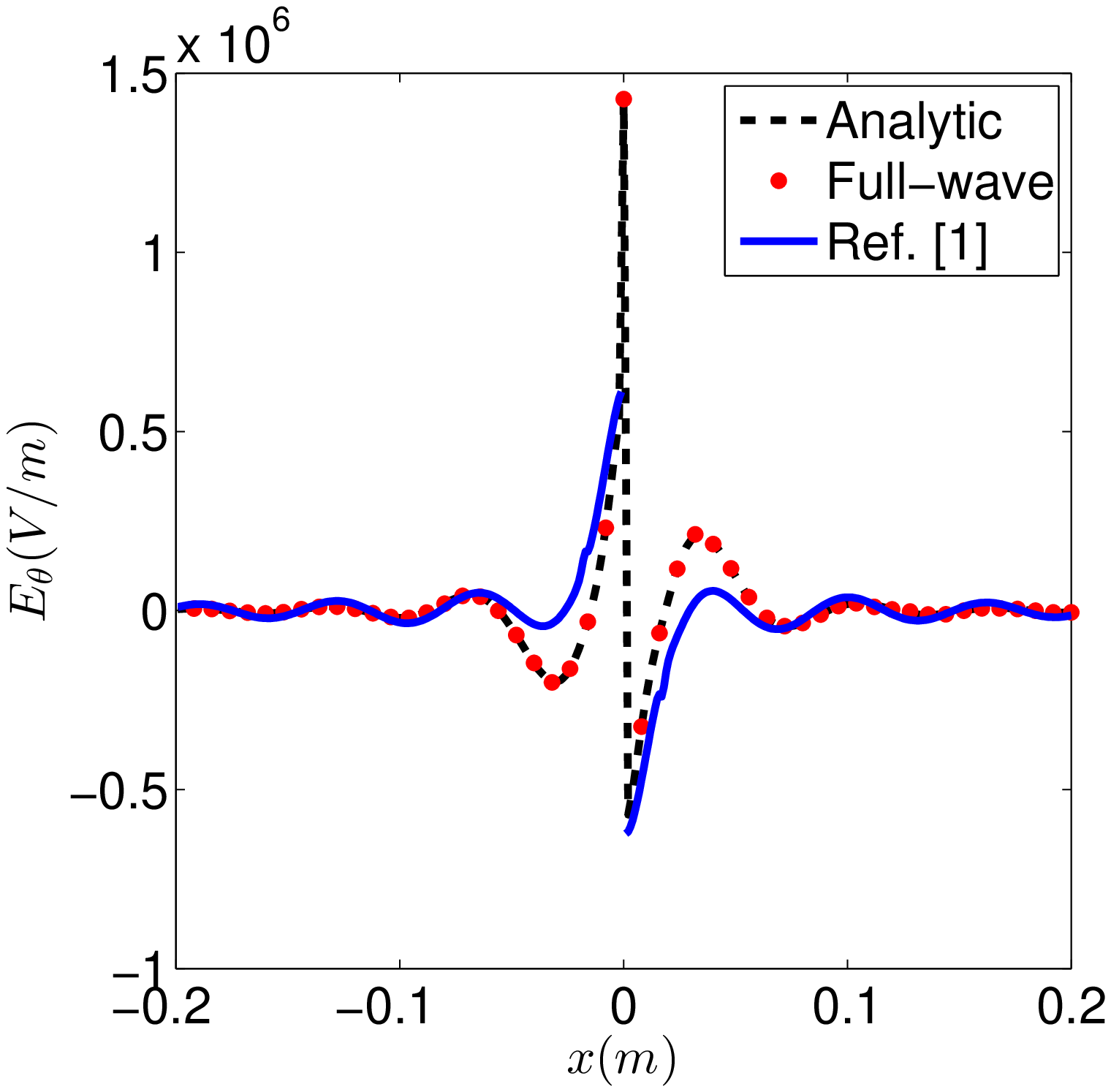}%
\label{fig:2c}}
\caption{Evaluation of the electric field radiated by a circular current loop. The plots depict a comparison between the proposed approach (black dashed line), full-wave simulation (red circles) and the approach in~\cite{Li:97} (blue solid line) for the~\protect\subref{fig:2a} radial, \protect\subref{fig:2b} azimuthal and~\protect\subref{fig:2c} polar components.}
\label{fig:02}
\end{figure}

Figure~\ref{fig:02} shows a plot of the evaluated electric field components using the different approaches. Very good agreement is observed between the results obtained analytically by our approach and those obtained by full-wave simulation for all electric field components. A good agreement with the results obtained by the approach in~\cite{Li:97} is observed in the exterior region for the radial and the polar components, whereas the field exhibits an unphysical discontinuity at the boundary between the exterior and the interior regions. The field values from~\cite{Li:97} in the interior region are not in agreement with the results of our approach neither with the full-wave simulation. Interestingly though, the azimuthal component does not exhibit any discontinuities at the transition between the interior and exterior regions, nevertheless, the computed field values greatly disagree with those obtained by our approach and the full-wave simulation in both regions. The artificial field discontinuities at the free-space boundary between the interior and exterior regions in~\cite{Li:97,Hamed:13} are likely attributed to the incompleteness of the field expansion set used.

\section{Conclusion}
We have presented an approach to evaluate the electromagnetic radiation by a circular loop carrying an arbitrary current. The electromagnetic fields are established by radially directed electric and magnetic Hertz potential functions. These choice of these potentials ensures that the electric potential generates the transverse electric field and the magnetic potential generates the transverse magnetic field, both with respect to the radial coordinate. The Hertz potentials are given in terms of an auxiliary Green's function. This auxiliary Green's function is determined using a polar transmission-line representation with eigenfunction expansion in the cross section using Fourier series representation in the azimuthal direction and Kontorovich-Lebedev transform in the radial direction. The field expressions are thus determined in a form of a series and an integral, and we show that the integral converges irrespective of the position of the loop or the observer. Moreover, we show that in the interior region, where the radial position of the observer is smaller than the radial position of the loop, an alternative representation of the integral in the form of a residue series sum exists.

The results obtained by the presented approach are compared to full-wave simulation results and an alternative approach presented elsewhere in the literature. The results show very good agreement with the full-wave simulation results, whereas the results of the alternative approach show unphysical field discontinuities in free-space. The presented approach thus provides a rigorous and physically correct method to determine the electromagnetic fields radiated by a circular current loop carrying arbitrary current.


%



\appendix[Hertz Potential and Field Expressions]

Substituting~\eqref{eq:1d_green_int} into~\eqref{eq:hertz} yields the following expression for Hertz potentials%
\begin{equation}
\label{eq:hertz_pe}
\begin{split}
r\Pi_{e} = \frac{k \pi}{2 \omega\epsilon r'} \sum_{m=0}^\infty \int_{-i\infty}^{i\infty} d \nu \mathcal{J} \mathcal{Z}^{(0,1)} \\
\times P_{\nu-\frac{1}{2}}^{-m}\left(\cos\left(\theta_<\right)\right) P_{\nu-\frac{1}{2}}^{-m}\left(-\cos\left(\theta_>\right)\right),
\end{split}
\end{equation}

\begin{equation}
\label{eq:hertz_ph}
\begin{split}
r\Pi_h^{<} = -\frac{k \pi}{2 r'} \sum_{m=0}^\infty \int_{-i\infty}^{i\infty} d \nu \mathcal{J} \sin\left(\theta'\right) \mathcal{Z}^{(0,0)} \Psi^{(0,-1)}, \\
r\Pi_h^{>} = \frac{k \pi}{2 r'} \sum_{m=0}^\infty \int_{-i\infty}^{i\infty} d \nu \mathcal{J} \sin\left(\theta'\right) \mathcal{Z}^{(0,0)} \Psi^{(-0,1)}, 
\end{split}
\end{equation}%
where $<|>$  respectively denote $\theta < \theta'$ and $\theta>\theta'$,%
\begin{align}
\mathcal{J} &= \frac{I_m e^{i m \phi} \Gamma\left(\nu+m-\frac{1}{2}\right)}{ \left[\nu+1\right] \Gamma\left(\nu-m-\frac{1}{2}\right) \sin\left(\pi \left[ \nu - \frac{1}{2} - m\right]\right)}, \nonumber \\
\Psi^{(\pm p, \mp q)} & = \frac{d^p P_{\nu-\frac{1}{2}}^{-m}\left(\pm \cos\left(\theta\right) \right)}{d \left[\cos\left(\theta\right)\right]^p} \frac{d^q P_{\nu - \frac{1}{2}}^{-m} \left(\mp \cos\left(\theta'\right)\right)}{d \left[\cos\left(\theta'\right)\right]^q}  ,\nonumber \\ 
\mathcal{Z}^{(p,q)} & = \frac{d^p}{d \left[k r\right]^p} j_{\nu-\frac{1}{2}}\left( k r \right) \frac{d^q}{d \left[k r'\right]^q} h_{\nu - \frac{1}{2}}^{(1)} \left(k r'\right), \nonumber 
\end{align}%
where $I_m$ are the Fourier expansion coefficients of the current. Inserting~\eqref{eq:hertz_pe} and~\eqref{eq:hertz_ph} into~\eqref{eq:field} yields the following expressions for the electric field%
\begin{equation}
\label{eq:er}
\begin{split}
E_r^{<} = \frac{k \pi}{2 \omega \epsilon r r'} \sum_{m=0}^\infty \int_{-i\infty}^{i\infty} d \nu m \mathcal{J} \mathcal{Z}^{(0,1)} \sin\left(\theta\right)^2 \\
\times \left[ \frac{m^2}{\sin\left(\theta\right)^4} \Psi^{(0,-0)}+ \frac{2 \cos\left(\theta\right)}{\sin\left(\theta\right)^2} \Psi^{(1,-0)} - \Psi^{(2,-0)} \right], \\
E_r^{>} = \frac{k \pi}{2 \omega \epsilon r r'} \sum_{m=0}^\infty \int_{-i\infty}^{i\infty} d \nu m \mathcal{J} \mathcal{Z}^{(0,1)} \sin\left(\theta\right)^2 \\
\times \left[ \frac{m^2}{\sin\left(\theta\right)^4} \Psi^{(-0,0)} - \frac{2 \cos\left(\theta\right)}{\sin\left(\theta\right)^2} \Psi^{(-1,0)} - \Psi^{(-2,0)} \right] ,
\end{split}
\end{equation}%
\begin{equation}
\label{eq:et}
\begin{split}
E_\theta^{<} = \frac{k \pi}{2 r'} \sum_{m=0}^\infty \int_{-i\infty}^{i\infty} d \nu m \mathcal{J} \frac{\sin\left(\theta\right)}{\sin\left(\theta'\right)} \\
\times \left\{ \omega \mu \frac{\sin\left(\theta'\right)^2}{\sin\left(\theta\right)^2} \mathcal{Z}^{(0,0)} \Psi^{(0,-1)} \right. \\
-\left. \frac{k}{\omega\epsilon r} \Psi^{(1,-0)} \left[ \mathcal{Z}^{(0,1)} + k r \mathcal{Z}^{(1,1)} \right] \right\} \\ 
E_\theta^{>} = \frac{k \pi}{2 r'} \sum_{m=0}^\infty \int_{-i\infty}^{i\infty} d \nu m \mathcal{J} \frac{\sin\left(\theta\right)}{\sin\left(\theta'\right)} \\
\times \left\{ - \omega \mu \frac{\sin\left(\theta'\right)^2}{\sin\left(\theta\right)^2} \mathcal{Z}^{(0,0)} \Psi^{(-0,1)} \right. \\
+\left. \frac{k}{\omega\epsilon r} \Psi^{(-1,0)} \left[ \mathcal{Z}^{(0,1)} + k r \mathcal{Z}^{(1,1)} \right] \right\},
\end{split}
\end{equation}%
\begin{equation}
\label{eq:ep}
\begin{split}
E_\phi^{<} = \frac{i k \pi}{2 r'} \sum_{m=0}^\infty \int_{-i\infty}^{i\infty} d \nu \frac{\mathcal{J}}{\sin\left(\theta\right) \sin\left(\theta'\right)} \\
\times \left\{ - \omega \mu \sin\left(\theta'\right)^2 \sin\left(\theta\right)^2 \mathcal{Z}^{(0,0)} \Psi^{(1,-1)} \right. \\
+\left. \frac{k m^2}{\omega\epsilon r} \Psi^{(0,-0)} \left[ \mathcal{Z}^{(0,1)} +  k r \mathcal{Z}^{(1,1)} \right] \right\} \\ 
E_\phi^{>} = \frac{i k \pi}{2 r'} \sum_{m=0}^\infty \int_{-i\infty}^{i\infty} d \nu \frac{\mathcal{J}}{\sin\left(\theta\right) \sin\left(\theta'\right)} \\
\times \left\{ - \omega \mu \sin\left(\theta'\right)^2 \sin\left(\theta\right) \mathcal{Z}^{(0,0)} \Psi^{(-1,1)} \right. \\
+\left. \frac{k m^2}{\omega\epsilon r} \Psi^{(-0,0)} \left[ \mathcal{Z}^{(0,1)} +  k r \mathcal{Z}^{(1,1)} \right] \right\},
\end{split}
\end{equation}%
and the following expressions for the magnetic field%
\begin{equation}
\label{eq:hr}
\begin{split}
H_r^{<} = -\frac{k \pi}{2 r r'} \sum_{m=0}^\infty \int_{-i\infty}^{i\infty} d \nu \mathcal{J} \mathcal{Z}^{(0,0)} \sin\left(\theta'\right) \sin\left(\theta\right)^2 \\
\times \left[ \frac{m^2}{\sin\left(\theta\right)^4} \Psi^{(0,-1)}+ \frac{2 \cos\left(\theta\right)}{\sin\left(\theta\right)^2} \Psi^{(1,-1)} - \Psi^{(2,-1)} \right], \\
H_r^{>} = \frac{k \pi}{2 r r'} \sum_{m=0}^\infty \int_{-i\infty}^{i\infty} d \nu \mathcal{J} \mathcal{Z}^{(0,0)} \sin\left(\theta'\right) \sin\left(\theta\right)^2 \\
\times \left[ \frac{m^2}{\sin\left(\theta\right)^4} \Psi^{(-0,1)} - \frac{2 \cos\left(\theta\right)}{\sin\left(\theta\right)^2} \Psi^{(-1,1)} - \Psi^{(-2,1)} \right] ,
\end{split}
\end{equation}%
\begin{equation}
\label{eq:ht}
\begin{split}
H_\theta^{<} = \frac{k \pi}{2 r r'} \sum_{m=0}^\infty \int_{-i\infty}^{i\infty} d \nu \mathcal{J} \sin\left(\theta\right) \sin\left(\theta'\right) \\
\times \left\{ k r \mathcal{Z}^{(1,0)} \Psi^{(1,-0)} +  \mathcal{Z}^{(0,0)} \Psi^{(1,-1)} \right. \\
+\left. \frac{m k r}{\sin\left(\theta\right)^2 \sin\left(\theta'\right)^2} \mathcal{Z}^{(0,1)} \Psi^{(0,-0)} \right\} \\ 
H_\theta^{>} = \frac{k \pi}{2 r r'} \sum_{m=0}^\infty \int_{-i\infty}^{i\infty} d \nu \mathcal{J} \sin\left(\theta\right) \sin\left(\theta'\right) \\
\times \left\{ k r \mathcal{Z}^{(1,0)} \Psi^{(-1,0)} +  \mathcal{Z}^{(0,0)} \Psi^{(-1,1)} \right. \\
+\left. \frac{m k r}{\sin\left(\theta\right)^2 \sin\left(\theta'\right)^2} \mathcal{Z}^{(0,1)} \Psi^{(-0,0)} \right\},
\end{split}
\end{equation}%
\begin{equation}
\label{eq:hp}
\begin{split}
H_\phi^{<} = -\frac{i k \pi}{2 r'} \sum_{m=0}^\infty \int_{-i\infty}^{i\infty} d \nu m \mathcal{J} \frac{\sin\left(\theta'\right)}{\sin\left(\theta\right)} \\
\times \left\{ k r \frac{\sin\left(\theta\right)^2}{\sin\left(\theta'\right)^2} \mathcal{Z}^{(0,1)} \Psi^{(1,-0)} \right. \\
+\left. \Psi^{(0,-1)} \left[ \mathcal{Z}^{(0,0)} +  k r \mathcal{Z}^{(1,0)} \right] \right\} \\ 
H_\phi^{>} = \frac{i k \pi}{2 r'} \sum_{m=0}^\infty \int_{-i\infty}^{i\infty} d \nu m \mathcal{J} \frac{\sin\left(\theta'\right)}{\sin\left(\theta\right)} \\
\times \left\{ k r \frac{\sin\left(\theta\right)^2}{\sin\left(\theta'\right)^2} \mathcal{Z}^{(0,1)} \Psi^{(-1,0)} \right. \\
+\left. \Psi^{(-0,1)} \left[ \mathcal{Z}^{(0,0)} +  k r \mathcal{Z}^{(1,0)} \right] \right\}.
\end{split}
\end{equation}

%
%

\ifCLASSOPTIONcaptionsoff
  \newpage
\fi

\end{document}